\documentclass[12pt]{article}      
\usepackage{graphicx}

\usepackage{amsthm}
\usepackage{amsmath,amsfonts,amssymb}
\newtheorem{theorem}{Theorem}
\newtheorem{definition}{Definition}

\newtheorem{corollary}{Corollary}
\newtheorem{lemma}{Lemma}

\usepackage{graphicx}
\usepackage{dcolumn}
\usepackage{bm}

\begin{document}


\title{Asymptotic behavior of mean fixation times in the Moran process with frequency-independent fitnesses}
\author{Ros\^angela A. Pires$^{1}$ \\Armando G. M. Neves$^{2}$
	\\
	\normalsize{$^{1}$Departamento de Matem\'atica, Universidade Federal de Minas Gerais}
	\\ 	\normalsize{r04assis@gmail.com}\\
	\normalsize{$^{2}$Departamento de Matem\'atica, Universidade Federal de Minas Gerais}	
	\\	\normalsize{aneves@mat.ufmg.br}
}

\date{\today}
\maketitle
\abstract{We derive asymptotic formulae in the limit when population size $N$ tends to infinity for mean fixation times (conditional and unconditional) in a population with two types of individuals, A and B, governed by the Moran process. We consider only the case in which the fitness of the two types do not depend on the population frequencies. Our results start with the important cases in which the initial condition is a single individual of any type, but we also consider the initial condition of a fraction $x\in (0,1)$ of A individuals, where $x$ is kept fixed and the total population size tends to infinity. In the cases covered by Antal and Scheuring (Bull Math Biol 68(8):1923–1944, 2006), i.e. conditional fixation times for a single individual of any type, it will turn out that our formulae are much more accurate than the ones they found. As quoted, our results include other situations not treated by them.}

\textbf{Keywords:} Markov chains, Asymptotic analysis, Birth-death processes

\section{Introduction} \label{secintro}
In this paper we will be concerned with the Moran process \cite{Moran} in its simpler setting: a population with fixed size equal to $N$ haploid individuals which reproduce asexually. There are two types of individuals, that we call A and B, distinguished e.g. by different alleles at a single locus. Time is considered to be discrete. At each time step two random processes happen: one individual produces a descendant having the same type as itself (\textit{no-mutation hypothesis}) and an individual dies, being replaced by the descendant of the reproducing individual. The reproduction and death lotteries are \textit{independent}. We consider the death lottery to be uniform, but the reproduction lottery is such that fitter individuals reproduce more frequently. In this paper we will suppose that the fitness of an individual depends only on its type A or B.

Taylor et al. \cite{taylor} extended the Moran process for two types of individuals to the Evolutionary Game Theory \cite{hofbauersigmund} setting in which fitnesses depend not only on individuals' types, but also on their population frequencies. The Moran process has been also extended to three or more types \cite{ferreirafixationthree} and to populations structured by graphs \cite{lieberman, Nowak, broomrychtar}. We will not treat any of these extensions in the present paper.

Another popular stochastic model for the same situation is the Wright-Fisher process \cite{fisher, wright}, in which at each time step the whole population is replaced by a new offspring generation, which composition depends on the parent generation. From a mathematical point of view, both Wright-Fisher and Moran processes are discrete-time Markov chains with a finite state space. Due to the no-mutation hypothesis, the states in which the population is either all-A, or all-B are \textit{absorbing}. All other states are \textit{transient}. Due to the existence of absorbing states, and also to the finiteness of the population, it can be shown \cite{allen, grinsteadsnell} that in both processes, after a long enough time, the population state will be, with probability 1, either all-A, or all-B. This mathematical phenomenon is called \textit{fixation}. Important quantities to be calculated are, for given initial composition of the population, the fixation probability of either type, the mean times for unconditional fixation, and for conditional fixation of either A or B individuals.

The transition probability from any transient state to any other state in the Wright-Fisher process is positive. On the contrary, the only non-zero transition probabilities in the Moran process are between states in which the number of A individuals differ at most by 1. As a consequence, for the Moran process there exist exact explicit formulae for quantities such as fixation probabilities \cite{Nowak}, see \eqref{probfixN}, and mean unconditional, see \eqref{formtigen}, or conditional fixation times \cite{Altrocktesis, AntalScheuring}, see \eqref{condmutA}, \eqref{condiA}. These formulae also hold when fitnesses depend on population frequencies. On the other hand, calculations of the same quantities in the Wright-Fisher model must rely on approximations, such as the diffusion approximation \cite{ewens}. For the extension of the Moran process to three or more types we also do not know exact formulae for fixation probabilities or times. The same for structured populations, except for some highly symmetric situations \cite{broomrychtar,hadjichrisanthou, monk}.

The possibility of some exact calculations is an advantage of the Moran process over Wright-Fisher's. Despite that, although exact and explicit, the alluded formulae for the Moran process are unwieldy in the sense that they do not immediately display qualitative and quantitative features of the fixation probabilities or times. As an example, they do not show how mean fixation times depend on the population size $N$. Some works have been devoted to developing simpler ways to understand them. Antal and Scheuring \cite{AntalScheuring} calculated -- in the game-theoretic Moran process -- the asymptotic behavior when the population size $N$ tends to infinity both for fixation probabilities and for conditional mean fixation times for a single individual of either type. de Souza et al. \cite{Souza} studied the shapes of the graphs for fixation probabilities in all scenarios of the game-theoretic Moran process, and also provided asymptotic formulae for the fixation probabilities for any population frequency of A individuals. Chalub and Souza \cite{Chalub2016} also have asymptotic formulae for the Moran process fixation probabilities, but restricted to what they call \textit{regular families of suitable birth-death processes}. This class includes the Moran process, but not in the strong-selection regime considered here and also in \cite{AntalScheuring} and \cite{Souza}.

This paper continues the work in \cite{Souza} and derives asymptotic formulae in the limit $N\rightarrow \infty$ for the mean fixation times, both conditional and unconditional, but restricted to the case of frequency-independent fitnesses.

The results in this paper are stronger in one sense than those of Antal and Scheuring \cite{AntalScheuring}, because they are more precise and mathematically fully rigorous, as will be further explained. We also include asymptotic formulae for the mean unconditional times and for conditional mean fixation times when the initial fraction of A individuals is a fixed $x \in (0,1)$. The part of the paper by Antal and Scheuring dealing with fixation times considers only mean conditional fixation times and only when there is a single individual of any type -- a mutant -- and the remainder of the population are individuals of the other type. But their results apply to the game-theoretic setting in which fitnesses depend on population frequencies. We avoid here these more complicated cases, postponing their treatment to another work.

The paper is organized as follows. In Section \ref{secMoranformulae} we define the Moran process for two types of individuals, introduce all the notation and terminology pertaining to it and display without proof the exact formulae on which the rest of the paper is based. After having defined the necessary terms, we also describe the results obtained in the paper. In Section \ref{sectimes} we consider the important case of initial condition consisting of a single individual of any type -- a mutant -- in a population of individuals of the other type and derive the asymptotic formulae for mean conditional fixation times and for mean absorption (i.e. unconditional fixation) times. We graphically compare the asymptotic results with the the exact formulae. In Section \ref{secgeninit} we fix the initial fraction of A individuals in the population and let population size $N$ tend to infinity. We derive asymptotic formulae for the mean conditional and unconditional fixation times. The results of these formulae are also compared with the exact ones. Some overall discussion is finally provided in Section \ref{secdisc}. Appendix \ref{secAux} is devoted to the statement of purely mathematical results and their proofs.

\section{Moran process: definitions and formulae} \label{secMoranformulae}
We start with the following definition \cite{Nowak}:
\begin{definition}\label{birthdeathprocess}
	A \emph{birth-death process} is a discrete-time Markov chain with finite state space $S=\{0,1,...,N\}$, in which states $0$ and $N$ are absorbing,  transitions between states $i$ and $j$ with $\lvert i-j\rvert \geq 2$ have probability 0 and, for $i \in \{1, 2, \dots, N-1\}$, transitions $i \to i \pm 1$ have non-zero probabilities. 
\end{definition}

It turns out that states $1, 2, \dots, N-1$ in a birth-death process are all transient and that fixation of the chain in one of the absorbing states occurs with total probability \cite{allen, grinsteadsnell}. If $X_n \in S$ denotes the state at time $n$, we define the relevant non-zero transition probabilities
\begin{equation}\label{ai}
	\alpha_i = P (X_{n+1} = i+1 \mid X_n = i)
\end{equation}
and
\begin{equation}\label{bi}
	\beta_i = P (X_{n+1} = i-1 \mid X_n = i) \; .
\end{equation}
As states $0$ and $N$ are absorbing, $\alpha_0 = \beta_N =0 $. 

We let $\pi_i$ denote the probability that a birth-death process is absorbed at state $N$, given that it started in state $i$. Of course, the corresponding probability for fixation at state 0 is $1-\pi_i$. Letting, for $i= 1, 2, \dots, N-1$,
\begin{equation}\label{ri}
	r_i = \frac{\alpha_i}{\beta_i}\; ,
\end{equation}
it can be shown, see e.g. chapter 6 in \cite{Nowak}, that
\begin{equation} \label{probfix1}
	\pi_1 = \frac{1}{1+ \sum_{j=1}^{N-1} \prod_{k=1} ^{j} r_{k} ^{-1}} \; ,
\end{equation}
and, for $i=2, \dots, N$
\begin{equation} \label{probfixN}
	\pi_i = \frac{1+ \sum_{j=1}^{i-1} \prod_{k=1} ^{j} r_{k} ^{-1}}{1+ \sum_{j=1}^{N-1} \prod_{k=1} ^{j} r_{k} ^{-1}} \; .
\end{equation}

We define now the mean times we will deal with in this paper:
\begin{definition}
	The \emph{mean absorption time} (or \emph{mean unconditional fixation time}) $t_i$ for a birth-death process is the expectation of the random time it takes to the Markov chain, starting at state $i$, to reach either of the absorbing states.
	
	The \emph{mean conditional fixation time} at state $N$ with initial condition $i$, denoted $t_i^N$, is the expectation of the random time it takes for the chain to go from state $i$ to state $N$, conditioned that it is absorbed at $N$. Analogously, $t_i^0$ is the mean conditional time for fixation at state $0$ starting from $i$.
\end{definition}

Of course, the mean absorption time and conditional mean fixation times are related by
\begin{equation} \label{relcondabs}
	t_i = \pi_i t_i^N+ (1-\pi_i) t_i^0 \;.
\end{equation}

As for fixation probabilities, exact formulae for the above defined times may be derived. Antal and Scheuring \cite{AntalScheuring} do the derivation for the mean conditional fixation times, but we will use different (but equivalent) formulae appearing in Altrock's thesis \cite{Altrocktesis}.
In this paper's notation, his formulae are
\begin{equation} \label{condmutA}
	t_1 ^N = \sum_{k=1}^{N-1} \left(\prod_{j=1}^k r_j \right) \sum_{n=1}^k \frac{\pi_n}{\alpha_n } \left(\prod_{\ell=1}^n r_{\ell}^{-1} \right)
\end{equation}
and, for $i=2,3, \dots, N-1$,
\begin{equation} \label{condiA}
	t_i ^N = \frac{1}{\pi_i }  \left[ \sum_{k=i}^{N-1}\left(\prod_{j=1}^k r_j \right)\sum_{n=1} ^{k} \frac{\pi_n}{\alpha_n } \left(\prod_{\ell=1}^n r_{\ell}^{-1}\right) \right]  -  \frac{1- \pi_i}{\pi_i} t_1 ^N \;.
\end{equation}

The corresponding formulae for $t_{N-1}^0$ and $t_i^0$ can be obtained from (\ref{condmutA}) and (\ref{condiA}) using an argument of \textit{duality} that we will use often and explain now. We define the dual birth-death process as another birth-death process in which birth and death are swapped, i.e. 
\[\overline{\alpha}_i = \beta_{N-i} \;\;\; \mathrm{and}\;\;\; \overline{\beta}_i = \alpha_{N-i}\;.\] 
Then
\begin{equation} \label{condiB}
	t_i^0= \overline{t}_{N-i}^N \;,
\end{equation}
valid for $i=1, 2, \dots, N-1$, where the dual time $\overline{t}_{i}^N$ is obtained by using (\ref{condiA}) or (\ref{condmutA}) with $\overline{r}_k=1/r_{N-k}$ replacing $r_k$.

With some work in \eqref{condiB} with $i=N-1$ and \eqref{condmutA}, it is possible to show \cite{AntalScheuring} that the mean conditional fixation times $t_1^N$ and $t_{N-1}^0$ are exactly equal. This is a consequence of a symmetry in conditional fixation times proved by Taylor et al. \cite{Taylor-Iwasa}. We may thus define 
\begin{equation}\label{tfixdefinicao}
	t_{fix} = t_{N-1} ^{0} =t_{1} ^{N} \;.
\end{equation}

The derivation of the formulae for the mean absorption times is similar. These formulae appear e.g. in \cite{Altrocktesis} and, translated to our notation, are
\begin{equation}\label{formt1gen}
	t_1 = 
	\pi_1 \sum_{j=1} ^{N-1}  \left(\prod_{k=1}^j r_k^{-1}\right)  \sum_{i=1} ^j \frac{1}{\alpha_i} \left(\prod_{\ell=1}^i r_{\ell}\right) 
	\;,
\end{equation}
and, for $i=2, 3, \dots, N-1$,
\begin{equation}\label{formtigen} 
	t_i =   \sum_{j=i}^{N-1} \left(\prod_{\ell=1}^j r_{\ell}^{-1}\right) \sum_{k=1}^j \frac{1}{\alpha_k} \left(\prod_{m=1}^k r_m\right)  -  t_1  \sum_{j=i} ^{N-1} \left(\prod_{\ell=1}^j r_{\ell}^{-1}\right) \; .
\end{equation}

The Moran process is a special case of birth-death process. The state $i \in S$ is identified as the number of A individuals in the population. Of course, the number of B individuals in a population at state $i$ is just $N-i$. In the more general case \cite{taylor, Nowak, Souza}, the fitnesses of A and B individuals may be calculated through a pay-off matrix and may depend on their population frequencies. In the simpler context of this paper, we suppose that type-A individuals have fitness $r>0$, whereas B individuals have fitness 1. Accordingly, parameter $r$ will be called \textit{the relative fitness of type-A individuals}. The Moran process probability for drawing an A individual for reproduction when the population is at state $i$ is defined as
\[ \frac{i r}{ir+N-i}\;.\] 
If $r>1$, A individuals are fitter than B. If $0<r<1$, it is the reverse. The important case $r=1$ is known as the \textit{neutral} Moran process.

With the above formula, the reproduction lottery favors fitter individuals, as remarked in Section \ref{secintro}. On the other hand, the death lottery will be considered uniform. Then, the probability of drawing a B individual for death in state $i$ is
\[\frac{N-i}{N}\;.\]
The transition $i \to i+1$ happens only if an A is drawn for reproduction and a B drawn for death. Its probability $\alpha_i$, defined in \eqref{ai}, is thus the product of the above probabilities:
\begin{equation}
	\label{aiMoran}
	\alpha_i = \frac{i (N-i)r}{N(ir +N-i)}\;.
\end{equation}
An analogous reasoning gives for the $i \to i-1$ transition
\begin{equation}
	\label{biMoran}
	\beta_i = \frac{i (N-i)}{N(ir +N-i)}\;.
\end{equation}

The ratio $r_i$ defined in \eqref{ri} becomes $r_i=r$, i.e. frequency-independent. We say that \eqref{aiMoran} and \eqref{biMoran} define the  Moran process with \textit{frequency-independent} fitnesses. The products and sums in \eqref{probfixN} can then be easily calculated and the fixation probability becomes
\begin{equation}  \label{piMoran}
	\pi_i=  \begin{cases}
		\frac{1-r^{-i}}{1 - r^{-N}}, \;\;\textrm{if $r \neq 1$}\\
		\frac{i}{N}, \;\;\textrm{if $r = 1$}
	\end{cases} \;.
\end{equation}

Frequency independence of $r_i$ causes many simplifications to arise in the formulae for the mean absorption and fixation times, too. First of all, we get from \eqref{aiMoran} that
\begin{equation} \label{alpha_n-ind}
	\frac{1}{\alpha_j} = N \left(\frac{1}{N-j} + \frac{1}{rj}\right) \; .
\end{equation}
Using this with \eqref{piMoran} in \eqref{formt1gen} we have, for $r \neq 1$, a simpler formula for the mean absorption time of an A mutant in a B population:
\begin{equation*}  
	t_1 = \frac{N}{r}\frac{1- r^{-1} }{1- r^{-N}} \sum_{k=1} ^{N-1} r^{-k} \sum_{j=1} ^k \left(\frac{r}{N-j} + \frac{1}{j}\right) r^{j} \;.
\end{equation*}
Reversing the summation order in the above and using the well-known formula for the sum of a finite geometric progression, we get, after some simple manipulations,
\begin{equation}\label{t1freqind}
	t_1 =  \frac{N(1+r)}{r (1- r^{-N})}  H_{N-1} -  \frac{N r^{-N}}{1- r^{-N}} \sum_{k=1}^{N-1} \left(\frac{1}{N-k}+\frac{1}{r k}\right) r^k   \;,
\end{equation}
where 
\begin{equation} \label{harmnum}
	H_n = \sum_{i=1}^n \frac{1}{i}
\end{equation}
are the \textit{harmonic numbers}.
Using duality, i.e. replacing $r$ by $r^{-1}$ in \eqref{t1freqind}, we obtain a formula for the mean absorption  time for a B mutant in an A population: 
\begin{equation}\label{tN-1freqind}
	t_{N-1} =  \frac{N(1+r)}{1- r^{N}}  H_{N-1} - \frac{N r^N}{1-r^{N}} \sum_{k=1}^{N-1} \left(\frac{1}{N-k}+\frac{r}{k}\right) r^{-k}  \;.
\end{equation}

For $t_i$, $t_{fix}$ and $t_i^N$, we can use again \eqref{piMoran} and \eqref{alpha_n-ind} respectively in \eqref{formtigen}, \eqref{condmutA} and \eqref{condiA}, reverse the summation order and use the formula for the sum of a geometric progression. We get similarly
\begin{align} \label{tiindfreq}
	t_i &=\frac{Nr}{r-1}\, \left[ r^{-i} \sum_{k=1}^{i-1} \left(\frac{1}{N-k}+\frac{1}{r k}\right) r^k - r^{-N} \sum_{k=1}^{N-1} \left(\frac{1}{N-k}+\frac{1}{r k}\right) r^k\right.\nonumber \\
	&\left. + H_{N-i}+ \frac{1}{r} (H_{N-1}-H_{i-1}) \right] - \frac{1-\pi_i}{\pi_1} \, t_1\;,
\end{align}
\begin{align}\label{tfixindfreq}
	t_{fix} &= \frac{N}{(r-1) (1 - r^{-N})} \left[\rule{0pt}{3ex}(1+ r)(1+ r^{-N})  H_{N-1} \right. \nonumber \\&\left.  
	-  r^{-(N-1)}  \sum_{k=1}^{N-1} \left(\frac{1}{N-k}+\frac{1}{r k}\right) r^k 
	- r \sum_{k=1}^{N-1} \left(\frac{1}{N-k}+\frac{1}{r k}\right) r^{-k}  \right] 
\end{align}
and
\begin{align}\label{tNiindfreq}
	t_{i}^N &= \frac{N r}{\pi_i(r-1) (1 - r^{-N})} \left[\rule{0pt}{3ex}r^{-N}(1+\frac{1}{r})H_{N-1}- r^{-i}(H_{N-1}-H_{N-i}+\frac{1}{r}H_{i-1})\right. \nonumber \\&  
	+H_{N-i}+ \frac{1}{r}(H_{N-1}-H_{i-1})- r^{-N}  \sum_{k=1}^{N-1} \left(\frac{1}{N-k}+\frac{1}{r k}\right) r^k \\
	&\left. +r^{-i} \sum_{k=1}^{i-1} \left(\frac{1}{N-k}+\frac{1}{r k}\right) r^k - \sum_{k=i}^{N-1} \left(\frac{1}{N-k}+\frac{1}{r k}\right) r^{-k}  \right] - \frac{1-\pi_i}{\pi_i}\, t_{fix}\;.\nonumber
\end{align}
We could have obtained a formula for $t_{i}^0$ similar to the above one by using duality, but we will not do it now. Later on, we will use another approach to derive a formula \eqref{t0i} for $t^0_i$.

Exact and simple formulae for all these quantities in the \textit{neutral} Moran process may be derived directly from  \eqref{condmutA}, \eqref{formt1gen}, \eqref{condiA} and \eqref{formtigen} setting all $r_j$ equal to 1. All products are trivially equal to 1 and sums can be written in explicit form. We state here, for completeness, the results for the \textit{neutral} Moran process. At first,
\begin{equation}  \label{neutralcond}
	t_{fix}= N(N-1)
\end{equation}
and
\begin{align} \label{neutralabsex}	t_1=t_{N-1}&= N H_{N-1} \\
	&= N \log N + N \, \gamma - \frac{1}{2} + O(\frac{1}{N}) \;,\label{neutralabsas}
\end{align}
where $\gamma \approx 0.577$ is the Euler-Mascheroni constant and the asymptotic approximation \eqref{neutralabsas} is consequence of a classical result due to Euler, see Lemma \ref{propde1sobren}, and simple manipulations with Taylor series. Next, we have
\begin{equation} \label{neutralti}
	t_i= N\left[(N-i)(H_{N-1}-H_{N-i-1})+i(H_{N-1}-H_i)\right]
\end{equation}
and
\begin{equation}  \label{neutralcondA}
	t^N_i= \frac{N(N-i)}{i} \left[N(H_{N-1}-H_{N-i})+1 \right]\;.
\end{equation}

The results of this paper in Section \ref{sectimes} are formulae such as \eqref{neutralcond} and \eqref{neutralabsas}, asymptotic in the limit $N \rightarrow \infty$, displaying explicitly how the mean conditional fixation and absorption times for the \textit{non-neutral} Moran process depend on the population size $N$ and on the relative fitness $r$ of A individuals, in the cases in which the initial condition of the population is \textit{a single individual of one type with $N-1$ individuals of the other type}. 

In Section \ref{secgeninit} we discuss the asymptotic behavior as $N \rightarrow \infty$ of the mean conditional absorption and fixation times with \textit{arbitrary initial conditions} of the population. We mean the asymptotic limit of the non-neutral analogs of  \eqref{neutralti} and \eqref{neutralcondA}, i.e. \eqref{tiindfreq} and \eqref{tNiindfreq}. We will use the exact result for the neutral case \eqref{neutralcondA} to illustrate one choice we made in this paper. 

There are two possible asymptotic limits when $N \rightarrow \infty$ for \eqref{neutralcondA}: we may either fix the number $i \in \mathbb{N}$ of A individuals in the initial condition, or fix the fraction $x \in (0,1)$ of A individuals in the initial condition.

If we fix $i$ and let $N \rightarrow \infty$, using \eqref{estimativa-h-n} and simple Taylor series, \eqref{neutralcondA} becomes
\begin{equation} \label{ifixed}
	t_i^N = N^2- \left(\frac{i}{2}+ \frac{1}{2i}\right) N+ O(1) \;.
\end{equation}
Instead, we may let $x = \frac{i}{N_0}$ for some $i, N_0 \in \mathbb{N}$, then take $N$ to be a multiple of $N_0$, so that $Nx$ is also integer. Letting $N \rightarrow \infty$, $N$ constrained to be a multiple of $N_0$ and using again \eqref{estimativa-h-n}, \eqref{neutralcondA} becomes
\begin{equation} \label{xfixed}
	t_{N x}^N = -\frac{(1-x)\log(1-x)}{x}\, N^2 - \frac{1}{2}\, N + O(1)\;.
\end{equation}

Equations \eqref{ifixed} and \eqref{xfixed} teach us different aspects about the mean  conditional fixation times for A individuals and both are interesting. In order not to overwhelm this paper, we chose in Section \ref{secgeninit} to pursuit only asymptotic formulas analog to \eqref{xfixed}, i.e. we fix the fraction $x$ of A individuals in the initial population and then make $N$  to infinity. The same choice was adopted for asymptotic formulae for fixation probabilities in \cite{Souza}.

\section{Mean conditional fixation and mean absorption times for a single mutant individual}\label{sectimes}
We start with the mean fixation time of a single individual -- a mutant -- of either type in a population with $N-1$ individuals of the other type \textit{conditioned} to fixation of the mutant's type. Although the fixation probabilities of a single A and of a single B individual may be quite different, their mean conditional fixation times are exactly equal \cite{Taylor-Iwasa}. This quantity, denoted as $t_{fix}$, is exactly given by \eqref{tfixindfreq} and its asymptotic behavior when the population size $N$ tends to infinity is calculated in the following result:
\begin{theorem}\label{theofix}
	Let $r$ be the relative fitness of A individuals in the non-neutral Moran process with frequency-independent fitnesses. 
	
	If $r > 1$,  
	\begin{align}\label{tfixrmaiorq1}
		t_{fix} = & N \left(\frac{r+1}{r-1} \right)\left[  \log N  + \gamma + \log( 1-r^{-1}) \right] +  \nonumber \\
		& + \frac{1}{2}\left(\frac{r+1}{r-1} \right)^2 +  O\left( \frac{1}{N} \right)  \; .
	\end{align}
	If $r \in (0,1)$,
	\begin{align}\label{tfixrmenorq1}
		t_{fix} = & N \left(\frac{r+1}{1-r} \right)\left[  \log N  + \gamma + \log( 1-r) \right] +  \nonumber \\
		&  + \frac{1}{2}\left(\frac{r+1}{1-r} \right)^2 +  O\left( \frac{1}{N} \right)  \; .
	\end{align}
\end{theorem}
\begin{proof}
	The right-hand side of \eqref{tfixindfreq} contains four sums and also the harmonic numbers $H_n$. All of these are estimated in the results in Appendix \ref{secAux}. We may obtain both \eqref{tfixrmaiorq1} and \eqref{tfixrmenorq1} by just using these results, taking the care to include in the remainders terms which are $O(1/N)$ or smaller.
	
	Alternatively, the formula for the case $r \in (0,1)$ may be obtained from the formula for $r>1$ by a duality argument: \eqref{tfixrmenorq1} is \eqref{tfixrmaiorq1} with $r$ replaced by $\overline{r}=1/r$.
\end{proof}

\begin{figure}\centering
	\includegraphics[width=\textwidth]{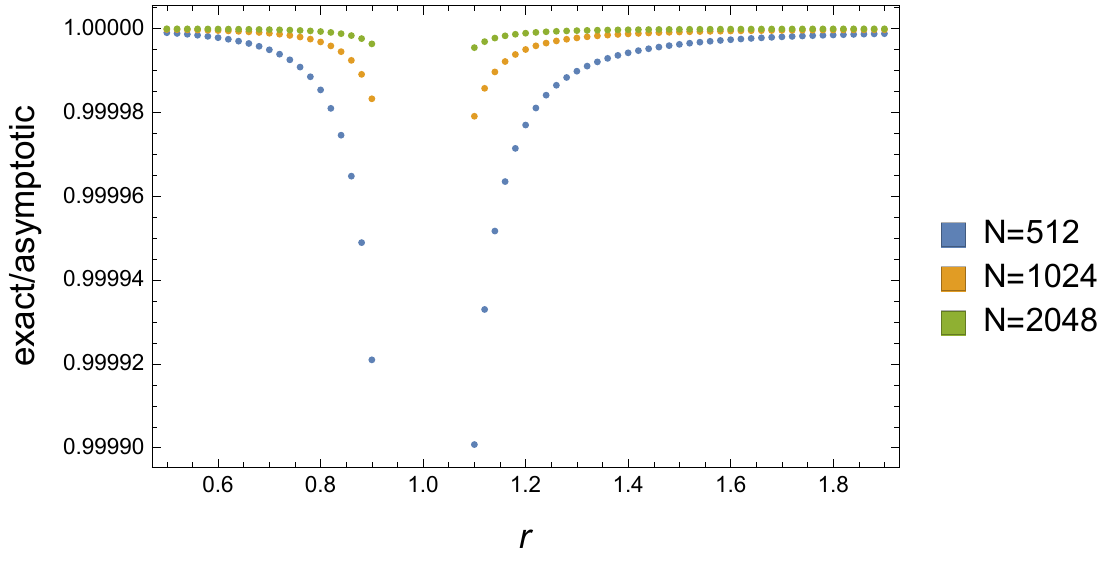}
	\caption{Ratio between $t_{fix}$ evaluated by the exact formula \eqref{tfixindfreq} and its asymptotic approximations given by \eqref{tfixrmaiorq1} and \eqref{tfixrmenorq1} as a function of $r$ for population values $N=512$, $N=1024$ and $N=2048$.}\label{figtfixr}
\end{figure}

Accuracy of the estimates in Theorem \ref{theofix} can be checked in Fig. \ref{figtfixr}, in which the vertical axis represents the ratio between the numerically evaluated $t_{fix}$ given by exact formula \eqref{tfixindfreq} and the asymptotic estimates \eqref{tfixrmaiorq1} and \eqref{tfixrmenorq1}. This figure should be compared with the analogous Fig. 5 in the paper by Antal and Scheuring \cite{AntalScheuring}. The scale of the vertical axes in our figures is of order $10^4$ times smaller than theirs, indicating a much larger accuracy.

We remind that our result is a special case of theirs, which was obtained for the more difficult case of the Moran process with frequency-dependent fitness, not treated in this paper. Despite that, their asymptotic estimate was carried out only to the dominant $O(N \log N)$ term, and thus, our estimate, containing $O(N)$ and $O(1)$ terms,  is much more precise.

The next situation we consider is the mean absorption time $t_1$ when a single A individual appears in a population with $N-1$ B individuals. It may happen that the final state is either all-A, or all-B and $t_1$ is the mean time it takes for one of them to happen. The result is given by
\begin{theorem} \label{theoabs}
	If $r > 1$, 
	\begin{align}\label{t1rmaiorq1}
		t_1  = &  \left(\frac{r + 1}{r} \right) N \log N  + \left[\left(\frac{r + 1}{r} \right) \gamma +  \log\left(1 - r^{-1} \right) \right] N   \nonumber \\
		& + \frac{1+r^2}{2r(r-1)} +  O\left( \frac{1}{N} \right)  \; .
	\end{align}
	If $ r \in (0,1)$, 
	\begin{equation}\label{t1rmenorq1}
		t_1 =  \left[ - \frac{\log(1-r)}{r} \right] N  + \frac{r}{1-r} +  O\left( \frac{1}{N} \right)  \; .
	\end{equation} 
	
\end{theorem}

\begin{proof}
	As in Theorem \ref{theofix}, it suffices to use the results in Appendix \ref{secAux}.
\end{proof}

The reader should notice that the absorption time when the population has a single A individual is $O(N \log N)$ if $r>1$, i.e. the mutant is fitter than the rest of the population, whereas it is only $O(N)$ if $0<r<1$, i.e. the mutant is less fit than the rest of the population. These are reasonable results, because there is an $O(1)$ probability, use \eqref{piMoran} with $i=1$, that the mutation is fixated if $r>1$, resulting in a longer absorption time of the same order as $t_{fix}$ when $r>1$. On the other hand, if $0<r<1$, the fixation probability for the mutant is $O(r^N)$, very small, and the most probable outcome is extinction of the mutant type. This will usually happen when the first mutant and its eventual offspring are drawn in the death lottery, which takes a time $O(N)$.

\begin{figure}\centering
	\includegraphics[width=\textwidth]{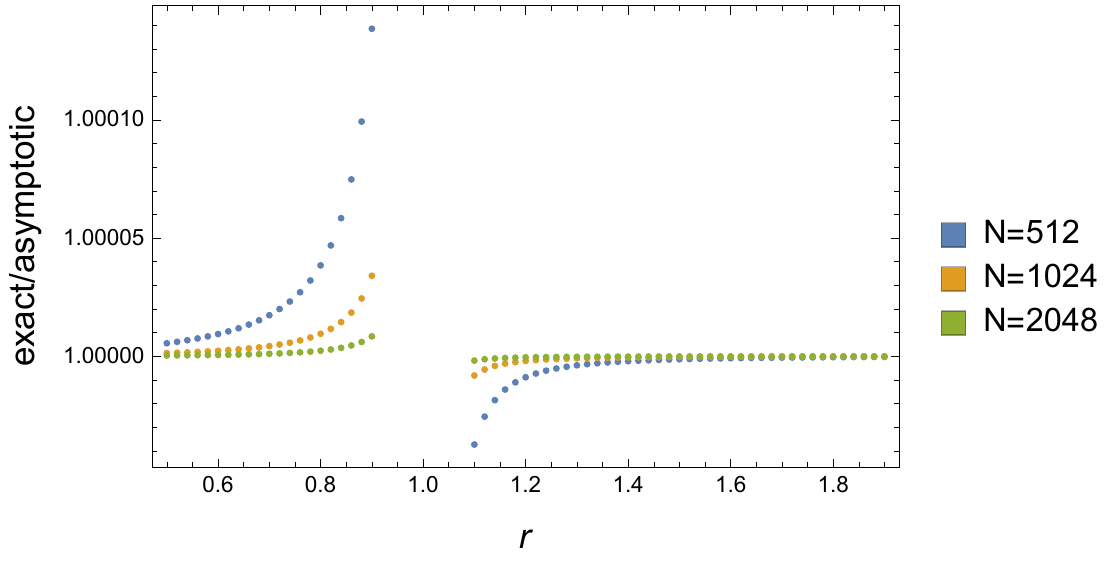}
	\caption{Ratio between $t_{1}$ evaluated by the exact formula \eqref{t1freqind} and its asymptotic approximations given by \eqref{t1rmaiorq1} and \eqref{t1rmenorq1} as a function of $r$ for population values $N=512$, $N=1024$ and $N=2048$.}\label{figt1}
\end{figure}

Fig. \ref{figt1} is analogous to Fig. \ref{figtfixr} and displays the high accuracy of the asymptotic results in Theorem \ref{theoabs}.

At last, we quote here the results for the mean absorption time $t_{N-1}$ for the situation of a single B mutant in a population with $N-1$ A individuals. Their proof is again a simple consequence of Appendix \ref{secAux}, or of Theorem \ref{theoabs} by noticing that a single B with fitness $r$ has the same mean absorption time as an A with fitness $\overline{r}=1/r$.

\begin{theorem} \label{theoabsN-1}
	If $r>1$,
	\begin{equation}\label{tN-1rmaiorq1}
		t_{N-1} =  \left[ - r \log(1-r^{-1})\right] N  + \frac{1}{r-1} +  O\left( \frac{1}{N} \right) \; .
	\end{equation}
	If $0<r<1$,
	\begin{align}\label{tN-1rmenorq1}
		t_{N-1} = & \left(r+ 1 \right) N \log N + \left[\left( r + 1 \right) \gamma +  \log\left(1 - r \right) \right]N \nonumber \\
		& - \frac{1+r^2}{2r(1-r)} +  O\left( \frac{1}{N} \right) \; .
	\end{align}
\end{theorem}

As the results for $t_{N-1}$ are obtainable from the analogous results for $t_1$, we found it unnecessary to illustrate their accuracy in a figure such as the previous ones.

\section{Mean fixation times for a population initially with a fixed fraction of A individuals}
\label{secgeninit}
As already anticipated by the end of Section \ref{secMoranformulae}, this section will deal with asymptotic formulae as $N \rightarrow \infty$ for the mean conditional fixation and mean absorption times in a population with an arbitrary initial condition given by a fixed \textit{fraction} $x$ of A individuals. Of course, the \textit{number} of A individuals in the initial population is $Nx$ and also tends to infinity in the desired asymptotic limit. 

\textbf{Warning:} In order that formulae \eqref{tiindfreq} and \eqref{tNiindfreq}, the starting points for our results here, make sense, $Nx$ must be integer. We take $x$ to be a rational number given by an irreducible fraction $p/q$ and take $N$ to be a multiple of $q$, $N=m q$, with $m \rightarrow \infty$. For simplicity, we will continue writing $N \rightarrow \infty$, but what we really mean is that $N=m q$, $m,q \in \mathbb{N}$, and $m \rightarrow \infty$.

The reader may notice that some terms such as $r^{-Nx}$ or $\frac{1}{(N (1-x))^2}$ will appear when producing the following results. The latter may be very large when $x$ is close to 1 and $N$ not too large. But, as the ratio between $\frac{1}{(N (1-x))^2}$ and $\frac{1}{N^2}$ is $\frac{1}{(1-x)^2}$, a term such as $\frac{1}{(N (1-x))^2}$ is correctly considered as $O(1/N^2)$. A similar situation holds for  $r^{-Nx}$. If $r>1$, but very close to 1, $x$ is close to 0 and $N$ is not too large, $r^{-Nx}$ may be almost as large as 1. But for fixed $x$ it decreases exponentially with $N$, if $r>1$, and can be safely included e.g. in term such as $O(1/N^2)$.

Formulae \eqref{tiindfreq} and \eqref{tNiindfreq} involve harmonic numbers and summations which had already appeared when asymptotically estimating $t_1$, $t_{N-1}$ and $t_{fix}$, but also other summations in which the number $i$ of A individuals appears as a summation index. Asymptotic estimates for these new summations, like for the other ones, are given in the Appendix \ref{secAux}.

Being simpler, we start with the result for the absorption times:
\begin{theorem}\label{theoti}
	Let $x$ be a fixed rational number in $(0,1)$.
	
	If $r>1$, then
	\begin{align}  \label{tnxr>1}
		t_{Nx} =& \frac{r}{r-1} N \log N + \frac{r}{r-1} N \left(\gamma+ \log(1-\frac{1}{r})+ \log(1-x) - \frac{1}{r} \log x\right) \nonumber \\
		&+ \frac{r+1}{2(r-1)^2} \left(\frac{r}{1-x}+\frac{1}{x}-1 \right) + O(\frac{1}{N}) \;.
	\end{align}
	
	If $0<r<1$,
	\begin{align}  \label{tnxr<1}
		t_{Nx} =& \frac{1}{1-r} N \log N + \frac{1}{1-r} N \left(\gamma+ \log(1-r)+ \log x - r \log (1-x)\right) \nonumber \\
		&+ \frac{r(1+r)}{2(1-r)^2} \left(\frac{1}{r x}+ \frac{1}{1-x} -1 \right) + O(\frac{1}{N}) \;.
	\end{align}
\end{theorem}

\begin{proof}
	We will describe how to get to \eqref{tnxr>1}. The other result may be obtained from \eqref{tnxr>1} by a duality argument: $t_i= \overline{t}_{N-i}$, where $\overline{t}_{N-i}$ is just $t_{N-i}$ with $r$ exchanged by $\overline{r}=1/r$. If $0<r<1$, then $\overline{r}>1$ and  $\overline{t}_{N-i}$ can be obtained by \eqref{tnxr>1}.
	
	So, we may now restrict to the $r>1$ case. We may write $i=Nx$ in \eqref{tiindfreq}. All terms in this formula are given appropriate estimates in the Appendix \ref{secAux}, including a term which had not appeared before, dealt with by Theorem \ref{theosumuptoi}.
	
	Using the appropriate results and including asymptotically negligible terms in a $O(1/N)$ remainder, we prove \eqref{tnxr>1}.
\end{proof}

\begin{figure}\centering
	\includegraphics[width=\textwidth]{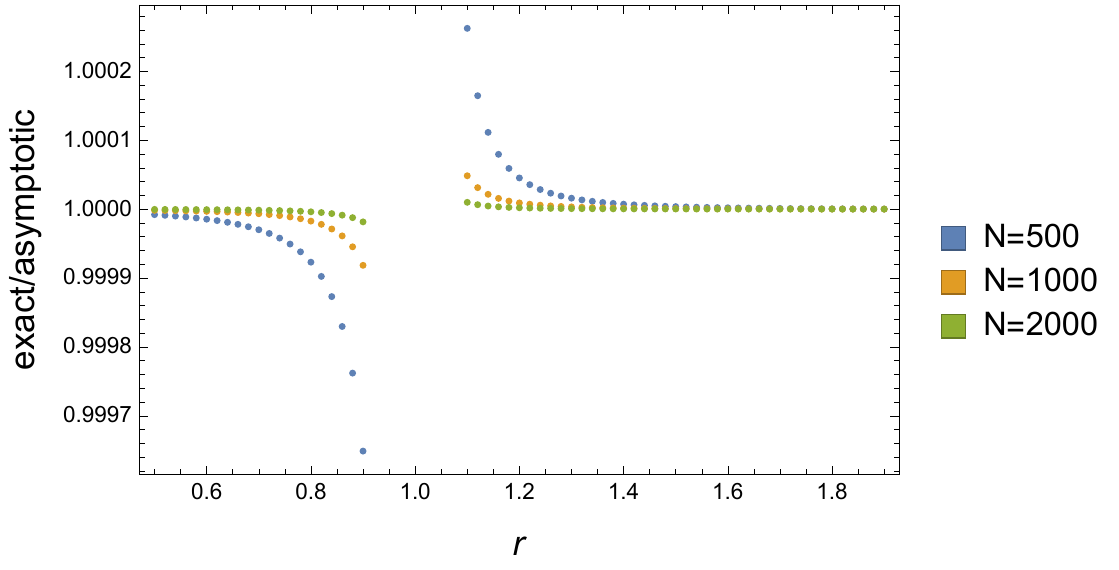}
	\caption{Ratio between $t_{0.4 N}$ evaluated by the exact formula \eqref{tiindfreq} and its asymptotic approximations given by \eqref{tnxr>1} and \eqref{tnxr<1} as a function of $r$ for population values $N=500$, $N=1000$ and $N=2000$.}\label{figtabsr}
\end{figure}

As in the previous section, we may check the accuracy of the above estimates. In Fig. \ref{figtabsr} we chose to fix an arbitrary fraction $x=0.4$ of A individuals in the population and show the ratio between the exact result and its asymptotic approximation. As seen, the asymptotic results for both $0<r<1$ and $r>1$ are quite accurate, although, as expected, they tend to get worse as $r$ approaches 1.

\begin{figure}\centering
	\includegraphics[width=\textwidth]{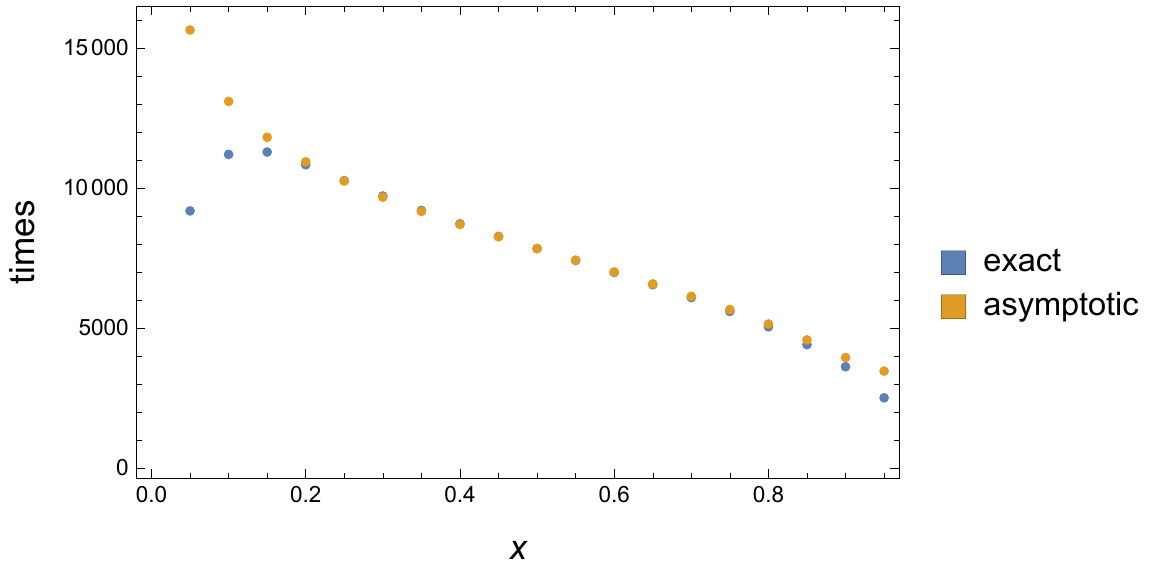}
	\caption{Comparison between $t_{200x}$ evaluated by the exact formula \eqref{tiindfreq} and its asymptotic approximation \eqref{tnxr>1} as a function of the fraction $x$ of A individuals in the population. Population size is $N=200$ and fitness of A individuals is $r=1.1$.}\label{figtabsx}
\end{figure}

In Fig. \ref{figtabsx} we fix instead the population size $N=200$ and fitness of A individuals $r=1.1$ and compare the exact value of $t_{N x}$ and its asymptotic approximations for several values of the fraction $x$ of A individuals. The difference between the exact and the asymptotic results is almost invisible if $x$ is not close to 0 or 1, where we expect that the approximation becomes worse. In fact, whereas the exact $t_{Nx}$ obeys boundary conditions of being equal to 0 when $x=0$ and $x=1$, the asymptotic approximations diverge at these limits. We have a sort of ``boundary layers" close to $x=0$ and $x=1$, with widths tending to 0 as $N \rightarrow \infty$, in which the asymptotic approximations do not match the boundary conditions.

We collect now all the results for conditional fixation times in the following
\begin{theorem}  \label{theoifixtimes}
	Let $x$ be a fixed rational number in $(0,1)$. If $r>1$, 
	\begin{align}  \label{tfcondAr>1}
		t^N_{Nx} =& \frac{r}{r-1} N \log N + \frac{r}{r-1} N \left(\gamma+ \log(1-\frac{1}{r})+ \log(1-x) - \frac{1}{r} \log x\right) \nonumber \\
		&+ \frac{r+1}{2(r-1)^2} \left(\frac{r}{1-x}+\frac{1}{x}-1 \right) + O(\frac{1}{N}) \;
	\end{align}
	and
	\begin{align}  \label{tfcondBr>1}
		t^0_{Nx} =& \frac{1}{r-1} N \log N + \frac{1}{r-1} N\, \left(\gamma+ \log(1-\frac{1}{r})-r \log(1-x) + \log x\right) \nonumber \\
		&+ \frac{r(r+1)}{2(r-1)^2} \left(\frac{1}{1-x}+\frac{1}{r x}-1 \right) + O(\frac{1}{N}) \;.
	\end{align}
	
	If $0<r<1$,
	\begin{align}  \label{tfcondAr<1}
		t^N_{Nx} =& \frac{r}{1-r} N \log N + \frac{r}{1-r} N \, \left(\gamma+ \log(1-r)-\frac{1}{r} \log x + \log (1-x)\right) \nonumber \\
		&+ \frac{1+r}{2(1-r)^2} \left(\frac{1}{x}+\frac{r}{1-x}-1 \right) + O(\frac{1}{N}) 
	\end{align}
	and
	\begin{align}  \label{fcondBr<1}
		t^0_{Nx} =& \frac{1}{1-r} N \log N + \frac{1}{1-r} N \left(\gamma+ \log(1-r)+ \log x - r \log (1-x)\right) \nonumber \\
		&+ \frac{r(1+r)}{2(1-r)^2} \left(\frac{1}{r x}+\frac{1}{1-x}-1 \right) + O(\frac{1}{N}) \;.
	\end{align}
\end{theorem}
\begin{proof}
	We replace $i$ by $Nx$ in \eqref{tNiindfreq}. If $r>1$ many of the terms in that equation, already estimated in Appendix \ref{secAux}, are negligibly small. For example,
	\[\frac{1-\pi_{Nx}}{\pi_{Nx}} = \frac{r^{-{Nx}}-r^{-N}}{1-r^{-Nx}}= O(r^{-Nx})\;.\]
	Using this with \eqref{tfixrmaiorq1} allows us to declare $\frac{1-\pi_{Nx}}{\pi_{Nx}} t_{fix}$ as negligible. Thus, stripping the negligible terms and simplifying, we get to
	\begin{align}
		t_{Nx}^N &= \frac{N r}{r-1} \left[  
		H_{N(1-x)}+ \frac{1}{r}(H_{N-1}-H_{Nx-1})- r^{-N}  \sum_{k=1}^{N-1} \left(\frac{1}{N-k}+\frac{1}{r k}\right) r^k \right.\\
		&\left. +r^{-Nx} \sum_{k=1}^{Nx-1} \left(\frac{1}{N-k}+\frac{1}{r k}\right) r^k - \sum_{k=N x}^{N-1} \left(\frac{1}{N-k}+\frac{1}{r k}\right) r^{-k}  \right] + O(r^{-Nx})\;.\nonumber
	\end{align}
	
	With the exception of the last sum, all the other terms had already appeared and were estimated. In Theorem \ref{theosumiuptoN} this last sum is asymptotically evaluated and results to be also negligible here. Estimate \eqref{tfcondAr>1} is thus obtained by using the results in the Appendix in the above formula. Observe that the right-hand sides of \eqref{tfcondAr>1} and \eqref{tnxr>1} are exactly the same. This is a consequence of \eqref{relcondabs} and the fact that when $r>1$, then  $\pi_{Nx}$ is $1-O(r^{-Nx})$, i.e. very close to 1.
	
	If next we try to obtain \eqref{tfcondAr<1} by directly following the same road we discover a large obstacle: if $0<r<1$, some terms in \eqref{tNiindfreq}, e.g. $r^{-N}(1+\frac{1}{r})H_{N-1}$, are huge, but $t^N_{Nx}$ cannot be larger than $t^N_0=t_{fix}$, which is only $O(N \log N)$. Cancellations of such huge terms among themselves must occur in order to get to the right formula. Our estimates in the Appendix \ref{secAux} are not accurate enough for that. 
	
	Instead, we use again \eqref{relcondabs}, from which we get
	\[t^0_i= \frac{1}{1-\pi_i} t_i - \frac{\pi_i}{1-\pi_i} t^N_i\;.\]
	We can use the above equation along with the exact results for $\pi_i$ \eqref{piMoran}, $t_i$ \eqref{tiindfreq} and $t^N_i$ \eqref{tNiindfreq} to get an exact formula for $t^0_i$:
	\begin{align}  \label{t0i}
		t^0_i=& \frac{N}{(1-\pi_i)(1-r^{-1})} \left\{ (1-\frac{1}{1-r^{-N}})\left[ r^{-i} \sum_{k=1}^{i-1} \left(\frac{1}{N-k}+\frac{1}{r k}\right)r^k  \right. \right.\nonumber\\
		-  &\left. \left. r^{-N} \sum_{k=1}^{N-1} \left(\frac{1}{N-k}+\frac{1}{r k}\right) r^k+ H_{N-i} + \frac{1}{r} (H_{N-1}-H_{i-1})  \right] \right. \nonumber\\
		-& \left. \frac{1}{1-r^{-N}} \left[ r^{-N}(1+ \frac{1}{r}) H_{N-1} - r^{-i} (H_{N-1} -H_{N-i} + \frac{1}{r} H_{i-1}) \right. \right. \\
		-& \left. \left. \sum_{k=i}^{N-1} \left(\frac{1}{N-k}+\frac{1}{r k}\right) r^{-k} \right] \right\}	+ t_{fix} - \frac{t_1}{\pi_1} \nonumber \;.
	\end{align}
	The above exact formula can be used to obtain \eqref{tfcondBr>1}. So, we consider $r>1$. Notice that the prefactor $\frac{1}{1-\pi_{Nx}}$ is very large: $r^{Nx}(1+O(r^{-N(1-x)}))$. On the other hand, $\frac{1}{1-r^{-N}} = 1+O(r^{-N})$, so that all terms in \eqref{t0i} which are multiplied by $1-\frac{1}{1-r^{-N}}$ are negligible. Stripping other negligible terms, it follows that
	\begin{align*}
		t^0_{Nx}=& \frac{N}{1-r^{-1}} r^{N x}  \left[ r^{-{N x}} (H_{N-1} -H_{N(1-x)} + \frac{1}{r} H_{N x-1})  \right] \\+& \sum_{k=N x}^{N-1} \left(\frac{1}{N-k}+\frac{1}{r k}\right) r^{-k}	+ t_{fix} - \frac{t_1}{\pi_1} \nonumber \;.
	\end{align*}
	Using the results of Appendix \ref{secAux}, Theorem \ref{theofix} and Theorem \ref{theoabs}, we prove \eqref{tfcondBr>1}.
	
	Having \eqref{tfcondAr>1} and \eqref{tfcondBr>1}, the remaining formulas can be easily obtained by duality \eqref{condiB}.
\end{proof}

\begin{figure}\centering
	\includegraphics[width=\textwidth]{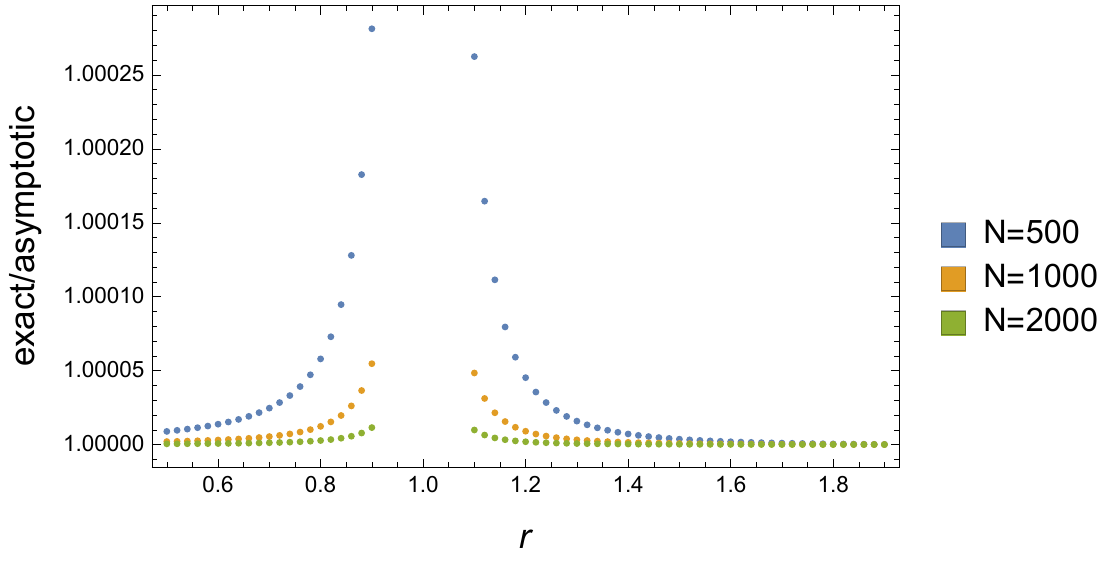}
	\caption{Ratio between $t^N_{0.4 N}$ evaluated by the exact formula \eqref{tNiindfreq} and its asymptotic approximations given by \eqref{tfcondAr>1} and \eqref{tfcondAr<1} as a function of $r$ for population values $N=500$, $N=1000$ and $N=2000$.}\label{figtfixAr}
\end{figure}

\begin{figure}\centering
	\includegraphics[width=\textwidth]{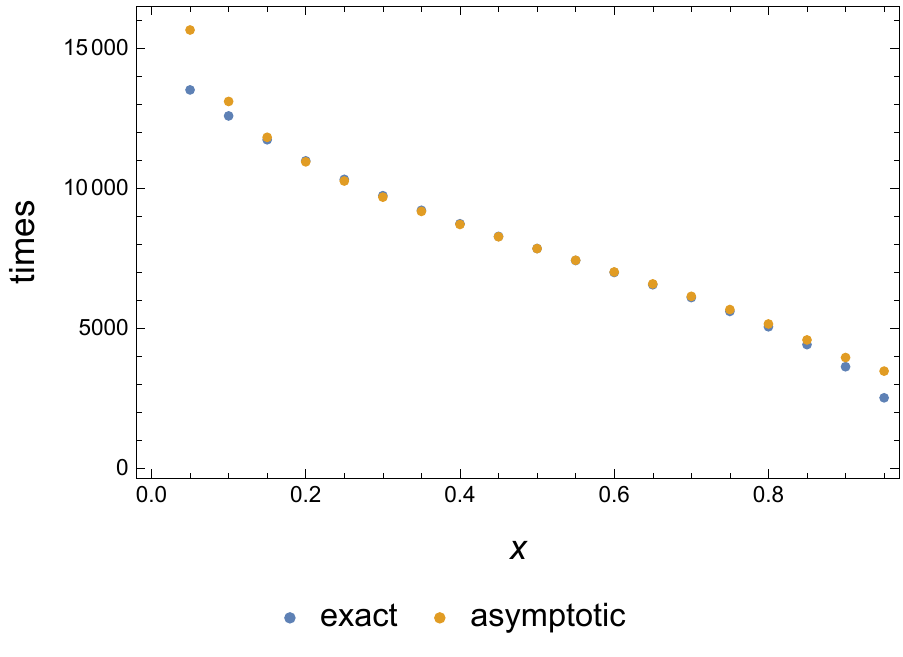}
	\caption{Comparison between $t^N_{200x}$ evaluated by the exact formula \eqref{tNiindfreq} and its asymptotic approximation \eqref{tfcondAr>1} as a function of the fraction $x$ of A individuals in the population. Population size is $N=200$ and fitness of A individuals is $r=1.1$.}\label{figtfixAx}
\end{figure}

The above results are illustrated in Figs. \ref{figtfixAr} and \ref{figtfixAx}, analogous to Figs. \ref{figtabsr} and \ref{figtabsx}, respectively.

\section{Discussion}\label{secdisc} 
The Moran process is an important stochastic model for the evolution of finite populations. In such populations, if mutations can be neglected, genetic traits will be fixated in the population. The Moran process with the \textit{no mutation hypothesis} exhibits the fixation phenomenon. But for how long can we expect that mutations at a certain trait will \textit{not} occur? Of course the larger population sizes are, the longer fixation times will be, whereas mutations will appear more frequently. We must be more quantitatively explicit on fixation times if we want to use a hypothesis of no-mutations. 

Although exact and explicit, formulae \eqref{probfixN}, \eqref{condmutA}, \eqref{condiA}, \eqref{formt1gen} and \eqref{formtigen} are not able to answer at a glance questions such as how will the mean absorption time change if population is doubled? Or how will the mean fixation time change if the fitness of A individuals increases by $10\%$?

The main reasons for such inadequacy is that all of these formulae involve sums of products in a complicated way. In particular, the population size $N$ appears as a summation limit.

Antal and Scheuring \cite{AntalScheuring} were first in answering such quantitative questions both for fixation probabilities and for mean fixation times in the limit when population size $N$ tends to infinity. Moreover, they did so in the more general setting in which fitnesses of the individuals depend on their population frequencies. 

Their approach involved transforming the product appearing in the alluded formulae in exponentials of sums of logarithms. This procedure was adopted again by Chalub and Souza \cite{Chalub2016}, which defined the \textit{fitness potential} further explored in \cite{Chalub2018}. One of us, with de Souza and Ferreira \cite{Souza}, also expanded the work of Antal and Scheuring on the asymptotic behavior of fixation probabilities when the population size $N$ tends to infinity. In doing so, we noticed the lack of complete mathematical rigor in the method of Antal and Scheuring. At a certain point, they approximate a sum by an integral, without taking into account the errors produced in doing so. Following \cite{Chalub2016}, these errors were called \textit{continuation errors} and we were able to show \cite{Souza} that they must be taken into account if accurate asymptotic formulae for the fixation probabilities are to be obtained.

If the fitnesses of types A and B do not depend on the population frequencies, the fixation probabilities are given by the simple formula \eqref{piMoran}, in which the asymptotic behavior when $N \rightarrow \infty$ is evident. The same cannot be said of formulae \eqref{t1freqind}, \eqref{tN-1freqind}, \eqref{tiindfreq}, \eqref{tfixindfreq} and \eqref{tNiindfreq} for the mean fixation times. This difference justifies the present work. 

In the simpler context of frequency-independent fitnesses, continuation errors and fitness potentials are not necessary, but still the above cited formulae do not explicitly display the asymptotic behavior for large $N$ of the mean fixation times on $N$, $r$ (relative fitness of A individuals) and $x$ (fraction of A individuals in the initial condition). Even numerically calculating with such exact formulae may be tricky. For example, if $0<r<1$, \eqref{tNiindfreq} contains terms multiplied by an $r^{-N}$ factor. These huge terms cancel among themselves, resulting in the $O(N \log N)$ result \eqref{tfcondAr<1}. In producing Fig. \ref{figtfixAx}, which relies also  numeric calculations with \eqref{tNiindfreq},  large numeric precision was necessary. Otherwise, \eqref{tNiindfreq} produces wrong results. The larger the $N$, more computing power is needed.

Not only we succeeded in this paper in obtaining results not present in \cite{AntalScheuring}, but our version for results present in both papers  is fully rigorous and much more accurate. This can be seen in comparing Figs. \ref{figtfixr} and \ref{figt1} with analogous figures in \cite{AntalScheuring}. Inclusion of sub-leading terms in the formulae of our Theorems \ref{theofix} to \ref{theoifixtimes} improved a lot the accuracy of the results. The reader should also notice that our formulae also contain the asymptotic order of the remainder terms.

One amazing fact we had already alluded to is that conditional fixation times for a single mutant individual, i.e $t^N_1$ and $t^0_{N-1}$, are exactly equal \cite{Taylor-Iwasa}, so that we use the same notation $t_{fix}$ for their common value. If $r \neq 1$ and $N$ is large, the fixation probabilities of the single A and B mutants are completely different: one of them is very close to 1 and the other decreases exponentially with $N$. Despite that huge probability difference, the corresponding mean conditional fixation times are exactly equal! 

Another similar astonishing result is obtained by comparing e.g. \eqref{tfcondAr>1} and \eqref{tfcondBr>1}, both valid for $r>1$: for any value of $x\in (0,1)$ and large enough $N$, B individuals, which are less fit, in average fixate faster than A individuals! That may not be surprising if the number of A individuals is small, i.e. $x$ is close to 0. But it holds even if $x$ is close to 1, if $N$ is large enough. As a numeric example using \eqref{tfcondAr>1} and \eqref{tfcondBr>1}, for $r=1.5$ and $x=0.8$, the mean conditional fixation time for A is approximately equal to the mean conditional fixation time for B forif $N \approx 10000$. For larger values of $N$, the Bs, even being only $20\%$ of the initial population and even having a very small fixation probability, in average fixate faster than the As. 

For the more general game-theoretic Moran process, in which fitnesses depend on the population frequencies, we should use formulae such as \eqref{condmutA}, \eqref{condiA}, \eqref{formt1gen} and \eqref{formtigen} instead of the simpler ones cited above. This will certainly involve consideration of continuation errors and will be left to a future work.

\section*{Acknowledgments}

This study was financed in part by the Coordena\c{c}\~ao de
Aperfei\c{c}oamento de Pessoal de N\'ivel Superior - Brasil (CAPES) -
Finance Code 001.

RAP received scholarships from CNPq (Conselho Nacional de Desenvolvimento Cient\'ifico e Tecnol\'ogico, Brazil) and CAPES.  AGMN is partially funded by Funda\c c\~ao de Amparo \`a Pesquisa do Estado de Minas Gerais (FAPEMIG), Brazil.

We thank the reviewer of a previous version for important suggestions that enriched the paper with new results.

\section*{Authorship}
Both authors contributed to the study conception and design. The first draft of the manuscript was written by Armando G M Neves. Both authors read and approved the final manuscript.

\section*{Data availability statement}
Data sharing not applicable to this article as no datasets were generated or analysed during the current study.

\section*{Competing interests}
The authors have no competing interests to declare that are relevant to the content of this article.

\begin{appendix}
	
	\section{Some auxiliary results and their proofs}\label{secAux}
	Our first result here is a classical consequence of the Euler-Maclaurin formula. It appears e.g. as formula (31) in \cite{apostol}.
	\begin{lemma}\label{propde1sobren}
		If $n \in \mathbb{N}$, then 
		\begin{equation}\label{estimativa-h-n}
			H_n =  \log n + \gamma + \frac{1}{2n} -\frac{1}{12 n^2} + O \left( \frac{1}{n^4} \right) \;,
		\end{equation}
		where $\gamma$ is the \textit{Euler-Mascheroni constant}. 
	\end{lemma}
	
	\begin{lemma}\label{lemmamaclog}
		For $z > 1$ we have
		\begin{equation}
			\sum_{k=1} ^{N-1} \frac{z^{-k}}{k} = - \log ( 1 - z^{-1}) + O\left( \frac{z^{-N}}{N} \right) \; .
		\end{equation}
	\end{lemma}
	\begin{proof}
		Notice that $\sum_{k=1} ^{\infty} \frac{x^k}{k}$ is just the Maclaurin series for $-\log(1-x)$, which converges if $0<x<1$. Thus, for $z>1$,
		\[\sum_{k=1}^{N-1} \frac{z^{-k}}{k} = - \log ( 1 - z^{-1}) - \sum_{k=N}^{\infty}\frac{z^{-k}}{k}\;. \]
		But
		\[\sum_{k=N}^{\infty}\frac{z^{-k}}{k} < \frac{1}{N} \, \sum_{k=N}^{\infty}z^{-k} =   \frac{1}{N} \, \frac{z^{-N}}{1-z^{-1}}\;. \]
	\end{proof}
	
	\begin{theorem} \label{sumrj/j}
		\begin{equation}  \label{estsumrj/j}
			\sum_{j=1} ^{N-1} \frac{r^j}{j}= \begin{cases}
				\frac{r^N}{N} \left[ \frac{1}{r-1} +O(\frac{1}{N})\right] \;, \;\;\textrm{if $r>1$}\\
				-\log(1-r) +O(\frac{r^N}{N})\;, \;\;\textrm{if $0<r<1$}
			\end{cases}\;.
		\end{equation}
	\end{theorem}
	\begin{proof}
		We start with the case $r>1$. Reversing the summation order, we get
		\begin{align*}\sum_{j=1} ^{N-1} \frac{r^j}{j} &= r^N\, \sum_{k=1}^{N-1}\frac{r^{-k}}{N-k} = \frac{r^N}{N}\, \sum_{k=1}^{N-1}\frac{r^{-k}}{1-\frac{k}{N}}\\
			&= \frac{r^N}{N}\left[\sum_{k=1}^{N-1}r^{-k} \,+\, \sum_{k=1}^{N-1}(\frac{1}{1-\frac{k}{N}}-1)\,r^{-k}\right] \;. \end{align*}
		
		The first sum in the above expression is equal to $\frac{r^{-1}-r^{-N}}{1-r^{-1}}=\frac{1}{r-1}+ O(r^{-N})$. In order to prove the estimate for $r>1$, we must show that the second sum is $O(1/N)$. To do that, we use the general identity
		\begin{equation}		\label{generalidentity}
			\frac{1}{1+y} = 1-y+ \frac{y^2}{1+y} \;,
		\end{equation}
		valid for any $y \neq -1$. Then,
		\[\frac{1}{1-\frac{k}{N}}-1 = \frac{k}{N}+ \left( \frac{k}{N}\right)^2 \,\frac{1}{1-\frac{k}{N}}\]
		and, 
		\[	\sum_{k=1}^{N-1}(\frac{1}{1-\frac{k}{N}}-1)\,r^{-k}= \frac{1}{N} \, \sum_{k=1}^{N-1} k r^{-k} + \frac{1}{N^2} \, \sum_{k=1}^{N-1} \frac{k^2}{1-\frac{k}{N}} r^{-k}\;.\]
		For $k \in \{1, 2, \dots, N-1\}$ we have $1- \frac{k}{N}\geq \frac{1}{N}$. We may also extend the summations up to $\infty$ and it results that
		\begin{align*}
			\sum_{k=1}^{N-1}(\frac{1}{1-\frac{k}{N}}-1)\,r^{-k} &< \frac{1}{N} \, \sum_{k=1}^{\infty} k\, r^{-k} + \frac{1}{N} \, \sum_{k=1}^{\infty} k^2\, r^{-k}
			\\ &= O(\frac{1}{N})\;.
		\end{align*}
		With this we prove the estimate for $\sum_{j=1} ^{N-1} \frac{r^j}{j}$ when $r>1$.
		
		The case $0<r<1$ is a simple consequence of Lemma \ref{lemmamaclog} with $z=r^{-1}$.
	\end{proof}
	
	\begin{corollary}\label{sumr-j/n-j}
		\begin{equation}
			\sum_{j=1} ^{N-1} \frac{r^{-j}}{j}= \begin{cases}
				-\log(1-r^{-1}) +O(\frac{r^{-N}}{N}) \;, \;\;\textrm{if $r>1$}\\
				\frac{r^{-N}}{N} \left[ \frac{r}{1-r} +O(\frac{1}{N})\right]\;, \;\;\textrm{if $0<r<1$}
			\end{cases}\;.
		\end{equation}
	\end{corollary}
	\begin{proof} It suffices to define $\overline{r}=1/r$, rewrite the sum in the left-hand side in terms of $\overline{r}$ and use Theorem \ref{sumrj/j} to produce the estimates.
	\end{proof}
	
	\begin{theorem} \label{sumrj/n-j}
		\begin{equation}
			\sum_{j=1}^{N-1} \frac{r^j}{N-j} = \begin{cases}
				-\log(1-r^{-1}) \, r^N+ O(\frac{1}{N})\;, \;\;\textrm{if $r>1$}\\
				\frac{1}{N} \, \frac{r}{1-r} \,+\, O(\frac{1}{N^2})	\;, \;\;\textrm{if $0<r<1$}
			\end{cases}\;.
		\end{equation}
	\end{theorem}
	\begin{proof}
		We replace the summation index $j$ by $k=N-j$, getting
		\[\sum_{j=1}^{N-1} \frac{r^j}{N-j} = r^N \, \sum_{k=1}^{N-1} \frac{r^{-k}}{k}\;.\]
		For $r>1$ the result follows by using Lemma \ref{lemmamaclog} with $z=r$. For $0<r<1$, we define $\overline{r}=1/r$ and use the first of the estimates in \eqref{estsumrj/j}.	
	\end{proof}
	
	For completeness, we state a simple consequence of Theorem \ref{sumrj/n-j} also appearing in the formulae for the mean fixation times:
	\begin{corollary}
		\begin{equation}
			\sum_{j=1} ^{N-1} \frac{r^{-j}}{N-j}= \begin{cases}
				\frac{1}{N} \frac{1}{r-1} +O(\frac{1}{N^2})\;, \;\;\textrm{if $r>1$}\\
				-\log(1-r) \, r^{-N}+ O(\frac{1}{N})\;, \;\;\textrm{if $0<r<1$}
			\end{cases}\;.
		\end{equation}
	\end{corollary}
	The proof goes as in Corollary \ref{sumr-j/n-j}.
	
	\begin{theorem}  \label{theosumuptoi}
		If $r>1$ and $i \in \{1,2,  \dots, N-1\}$,
		\begin{align} \label{somtni}
			&\sum_{k=1}^{i-1} \frac{r^k}{N-k} = r^i \, \left\{\frac{1}{N-i} \, \frac{r^{-1}-r^{-i}}{1-r^{-1}} \right. \nonumber \\  &\left. -\frac{1}{(N-i)^2} \left[\frac{r^{-1}}{(1-r^{-1})^2} -
			\frac{r^{-i}}{(1-r^{-1})^2}(\frac{1}{r}+(1-\frac{1}{r})i) \right]+ O\left(\frac{1}{(N-i)^3}\right)\right\}\;.
		\end{align}	
	\end{theorem}
	\begin{proof}
		First, we define $j=i-k$, so that the summation becomes
		$r^{i}\, \sum_{j=1}^{i-1} \frac{r^{-j}}{N-i+j}$.
		We write then
		\begin{align*}
			\frac{1}{N-i+j} &= \frac{1}{N-i} \, \frac{1}{1+ \frac{j}{N-i}}\nonumber \\
			&= \frac{1}{N-i} \, \left[1- \frac{j}{N-i} + \frac{j^2}{(N-i)^2}\frac{1}{1+ \frac{j}{N-i}}\right] \;,
		\end{align*}
		where we used again \eqref{generalidentity} in the last passage.	Substituting this in the summation we get
		\begin{align*}
			&\sum_{k=1}^{i-1} \frac{r^k}{N-k}  = \frac{1}{N-i}\, \left[\sum_{j=1}^{i-1} r^{-j}-\frac{1}{N-i}\, \sum_{j=1}^{i-1} j\, r^{-j} + \frac{1}{(N-i)^2} \sum_{j=1}^{i-1} \frac{j^2}{1+ \frac{j}{N-i}}r^{-j} \right] \;.
		\end{align*}
		The first sum inside the square brackets is
		\[\sum_{j=1}^{i-1} r^{-j} = \frac{r^{-1}-r^{-i}}{1-r^{-1}} \;.\]
		The second may be calculated from the first:
		\[\sum_{j=1}^{i-1} j\, r^{-j} = -r \frac{d}{dr} \, \sum_{j=1}^{i-1} r^{-j}\;.\]
		The third sum may be bounded as
		\[\sum_{j=1}^{i-1} \frac{j^2}{1+ \frac{j}{N-i}}r^{-j} < \sum_{j=1}^{\infty} j^2 \,r^{-j} = O(1)\;.\]
		Substituting the above three results, we get to the right-hand side of \eqref{somtni}.
	\end{proof}
	
	\begin{theorem}  \label{theosumiuptoN}
		If $r>1$ and $i \in \{1,2,  \dots, N-1\}$,
		\begin{align} \label{first}
			&\sum_{k=i}^{N-1} \frac{r^{-k}}{N-k} = \frac{r^{-i}}{N-i} \left[\frac{1}{1-r^{-1}}+ O(\frac{1}{N-i})\right]
		\end{align}	
		and
		\begin{align} \label{second}
			&\sum_{k=i}^{N-1} \frac{r^{-k}}{k} =	\frac{r^{-i}}{i} \left[\frac{1}{1-r^{-1}}+ O(\frac{1}{i})\right]\;.
		\end{align}	
	\end{theorem}
	\begin{proof}
		We use the new summation index $j=k-i$ to transform the sum in the left-hand side of \eqref{first} into
		\[\frac{r^{-i}}{N-i} \sum_{j=0}^{N-1-i} \frac{r^{-j}}{1-\frac{j}{N-i}}\;.\]
		
		We can then use \eqref{generalidentity} to write
		\begin{align} 
			&\sum_{k=i}^{N-1} \frac{r^{-k}}{N-k} = \frac{r^{-i}}{N-i} \left[\sum_{j-0}^{N-1-i} r^{-j} + \frac{1}{N-i}\sum_{j-0}^{N-1-i} j \,r^{-j} \right. \nonumber \\ 
			+& \left. \frac{1}{(N-i)^2}\sum_{j-0}^{N-1-i} \frac{j^2}{1-\frac{j}{N-i}} \,r^{-j}\right]\;.
		\end{align}
		The first sum is $\frac{1}{1-r^{-1}}+O(r^{-(N-i)})$ and the second is $O(1)$. In the third sum, we may bound $\frac{1}{1-\frac{j}{N-i}}$ simply by 1 and show it is $O(1)$. From these estimates, \eqref{first} follows.
		
		Almost the same tricks can be used to prove \eqref{second}.
	\end{proof}

\end{appendix}



\end{document}